# Methods of Estimation for the Three-Parameter Reflected Weibull Distribution

*Fateme Maleki jebeli [a]*, *Einolah Deiri [b,\*]*

[a]Department of Statistics, Marvdasht Branch, Islamic Azad University, Marvdasht, Iran

Email: f.maleki7351@yahoo.com

[b]Department of Statistics, Qaemshahr Branch, Islamic Azad University, Qaemshahr, Iran

Email: e.deiri53@qaemiau.ac.ir

**Abstract**

In this paper, we propose methods for the estimation of parameters for the three-parameter Reflected Weibull ($RW$) distribution. The Moment estimator ($MME$), Maximum likelihood estimator ($MLE$) and Location and Scale Parameters free maximum likelihood estimator ($LSPFE$). The $LSPFE$ is based on a data transformation, which avoids the problem of unbounded likelihood estimator. Through Mont Carlo simulations, we further show that the $LSPFE$ performs better than $MME$ and $MLE$ in terms of bias and root mean squared error ($RMSE$). Finally, two examples based on real data sets are presented to illustrate methods.

**Introduction**

The Weibull distribution, first presented by Weibull [17], is the most widely used distribution in reliability and lifetime studies. The cumulative distribution function ($CDF$) and probability density function ($PDF$) of the three-parameter Weibull distribution are given by

$$F(x; \delta, \beta, \gamma) = 1 - \exp\left[-\left(\frac{x-\gamma}{\beta}\right)^{\delta}\right] \quad (1)$$

and

$$f(x; \delta, \beta, \gamma) = \frac{\delta}{\beta}\left(\frac{x-\gamma}{\beta}\right)^{\delta-1} exp\left[-\left(\frac{x-\gamma}{\beta}\right)^{\delta}\right] \quad (2)$$

for $\delta > 0$, $\beta > 0$ and $\gamma < x$, for example see, Johnson et al. [8].

If X has the Weibull distribution with $CDF$ and $PDF$ given by (1) and (2) then $-X$ is said to have the RW distribution. The $CDF$ and $PDF$ for the three-parameter $RW$ are given by

$$F(x; \delta, \beta, \gamma) = 1 - \exp\left[-\left(\frac{\gamma-x}{\beta}\right)^{\delta}\right] \quad (3)$$

and

$$f(x; \delta, \beta, \gamma) = \frac{\delta}{\beta}\left(\frac{\gamma-x}{\beta}\right)^{\delta-1} exp\left[-\left(\frac{\gamma-x}{\beta}\right)^{\delta}\right] \quad (4)$$

for $\delta > 0, \beta > 0$ and $x < \gamma$. The associated mean $E(x) = \gamma - \beta \Gamma\left(\frac{1}{\gamma} + 1\right)$; where $\Gamma(x)$ is gamma function

$$\Gamma(x) = \int_0^\infty t^{x-1} e^{-x} dx.$$

This distribution, first presented by Cohen[4].
For $\delta \leq 1$, the distribution is J-shape, for $\delta > 1$, the $RW$ distribution becomes bell-shape.

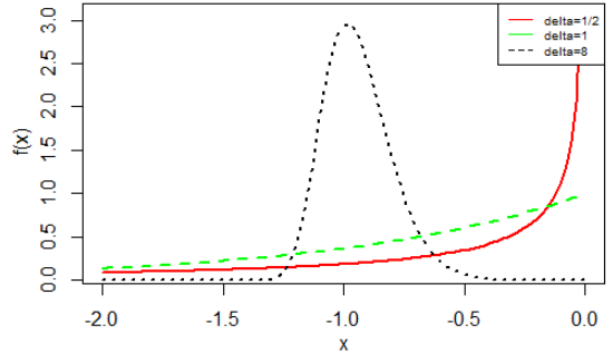

**Figure1.** The density function of the three parameter $RW$ distribution for different choices of $\delta$ where $\beta = 1, \gamma = 0$.

As Cohen[4] has said some readers may recognize the $RW$ distribution of largest values, or the Fisher-Tippet type III distribution of largest values as discussed by Gumbel[7].
As Lai [9] has said strictly speaking, the $RW$ is not suitable for reliability modeling unless $\gamma > 0$ and $\left(\frac{\gamma}{\beta}\right)^\delta \geq 9$.

The RW distribution is suitable for ductile strength, you can see Nadarajah and Kotz.[12].
In this paper, we propose three method of estimation of parameters of the three-parameter $RW$ distribution. $MME$ and $MLE$ that discussed with many of authors for a lot of distributions. As Chen and Amin [3] said $MLE$ does not always give satisfactory estimates of parameters for certain three-parameter distributions where the density is positive only to right of a shifted origin, $\gamma$, this being of the unknown parameters. for example in the Lognormal, Gamma distribution and Weibull model with three parameters the critical difficulty is that there are paths in the parameter space, with $\gamma$ tending to the smallest observation, along which the likelihood becomes infinite.
Griffths [6] suggested a method for estimation parameters of the three-parameter Lognormal distribution. Lawless [10] have all given detailed of descriptions of various methods of parameter estimation of the three-parameter Weibull distribution.

As Nagatsuka and Balakrishnan [13] said since there are estimation that are uniqueness but they are useless from an inferential point of views and consistency is one of the most fundamental properties to show that statistics are suitable as estimators of unknown parameters and as Nagatsuka et al. [14] suggested we will say $LSPFE$s.

The rest of this article is organized as follows. In section 1, we present the $MLE$. In section 2, we present the $MME$. In section 3, we present the $LSPFE$, as a new method for estimation of parameters of the three-parameter $RW$ distribution. In section 4, we show that the $LSPFE$ performs well compared to some other prominent methods, we will simulate and use of bias and $RMSE$. In section 5, two real life data sets are used as examples to illustrate the methods of estimation. In section 6 we will write some concluding remarks.

## 1. Maximum likelihood estimation

Let $X_i, i = 1, 2, \ldots n$, be a random variable distributed as (3) with the vector of parameters $(\delta, \beta, \gamma)$. We now determine the $MLE$s of the parameters of the three-parameter $RW$ distribution. Let $x_1, x_2, \ldots, x_n$ be observed values of a random sample size $n$ from the three-parameter $RW$ distribution. The log-likelihood function for the vector of parameters can be written as

$$l(\delta, \beta, \gamma) = n\log\delta - n\log\beta + (\delta - 1) \sum_{i=1}^{n} \log\left(\frac{\gamma - x_i}{\beta}\right) - \sum_{i=1}^{n} \left(\frac{\gamma - x_i}{\beta}\right)^{\delta}$$

When $x_i < \gamma, i = 1, 2, \ldots n$, and

$$l(\delta) = \frac{\partial l(\delta, \beta, \gamma)}{\partial \delta} = \frac{n}{\delta} + \sum_{i=1}^{n} \log\left(\frac{\gamma - x_i}{\beta}\right) - \sum_{i=1}^{n} \left(\frac{\gamma - x_i}{\beta}\right)^{\delta} \log\left(\frac{\gamma - x_i}{\beta}\right),$$

$$l(\beta) = \frac{\partial l(\delta, \beta, \gamma)}{\partial \beta} = \frac{-n\delta}{\beta} + \frac{\delta}{\beta} \sum_{i=1}^{n} \left(\frac{\gamma - x_i}{\beta}\right)^{\delta+1}$$

$$l(\gamma) = \frac{\partial l(\delta, \beta, \gamma)}{\partial \gamma} = (\delta - 1) \sum_{i=1}^{n} \left(\frac{1}{\gamma - x_i}\right) - \frac{\delta}{\beta} \sum_{i=1}^{n} \left(\frac{\gamma - x_i}{\beta}\right)^{\delta-1}$$

As we know, The $MLE$ of $\gamma$ is more than $X_{(1)}$, where $X_{(i)}$ denotes $i$-th order statistic. The $MLE$ of $\delta$, $\beta$ and $\gamma$ are obtained by solving the non-linear equations $l(\delta) = 0$, $l(\beta) = 0$ and $l(\gamma) = 0$.

## 2. Moments Estimation

We know

$$E(\gamma - X)^k = \int_{-\infty}^{\gamma} \left(\frac{\delta}{\beta}\right) (\gamma - x)^k \left(\frac{\gamma - x}{\beta}\right)^{\delta-1} \exp\left[-\left(\frac{\gamma - x}{\beta}\right)^{\delta}\right] dx$$

If $\left(\frac{\gamma - x}{\beta}\right)^{\delta} = u$ then

$$E(\gamma - X)^k = \beta^k \, \Gamma\left(\frac{k}{\delta} + 1\right).$$

We can write

$$E(X) = \gamma - \beta\Gamma\left(\frac{1}{\delta} + 1\right),$$
$$E(X^2) = 2\gamma E(X) - \gamma^2 + \beta^2 \Gamma\left(\frac{2}{\delta} + 1\right),$$
$$E(X^3) = \gamma^3 - 3\gamma^2 E(X) + 3\gamma E(X^2) - \beta^3 \Gamma\left(\frac{3}{\delta} + 1\right)$$

with replace $E(X), E(X^2)$ and $E(X^3)$ with $\bar{x} = \frac{1}{n}\sum_{i=1}^{n} x_i$, $\overline{x^2} = \frac{1}{n}\sum_{i=1}^{n} x_i^2$ and $\overline{x^3} = \frac{1}{n}\sum_{i=1}^{n} x_i^3$, we simply obtain the moment estimates of $\delta, \beta$ and $\gamma$.

## 3. Location and scale parameters free maximum likelihood Estimation

It is well known that the reliability conditions are not satisfied for the $MLE$ for every distribution with three-parameter then some authors suggested a new method for this problem. For example;

Nagatsuka and Balakrishnan [13] studied about methods of estimation for three-parameter Inverse Gaussian distribution. Nagatsuka et al. [15] studied about methods of estimation for three-parameter Gamma distribution. Nagatsuka et al. [14] studied about methods of estimation for three-parameter Weibull distribution and suggested that authors can be study about another distributions such as the three-parameter $RW$ distribution. Then in this paper we want studied this new method for estimation of parameters for this distribution and compared with the $MLE$ and $MME$. In this section we will say about this new method.

### 3.1. Estimation of the shape parameter

In this section, we describe a new method of estimation of the parameters of the three - parameters RW distribution and discuss some of properties. Let $X_1, X_2, \ldots, X_n$ be $n$ independent random variables from the three- parameter $RW$ distribution with common $CDF$ (3). Throughout the paper, we assume that the following two conditions hold:

**Assumption** 1. The sample size $n$ is greater than 2.
**Assumption** 2. With $X_i \neq X_j$ probability 1, for some $i \neq j$.

These conditions are very natural, which are required for all existing methods of estimation for three-parameter distributions. We first consider some statistics depending on only one parameter a before presenting the method of estimation. For $i = 1,2,\ldots,n$, let $X_{(i)}$ denote the order statistics among $X_1, X_2, \ldots, X_n$. Then, we consider the following statistics:

$$W_{(i)} = \frac{X_{(i)} - X_{(1)}}{X_{(n)} - X_{(1)}}, \quad i = 1,2,\ldots,n \tag{5}$$

It is easy to see that $W'_{(i)}$s do not depend on $\gamma$ and $\beta$, but depend only on $\delta$. Statistics similar to those in (5) have been considered by Nagatsuka and Kamakura [16] for the model presented by Castillo and Hadi [2]. Observe that $W_{(1)}$ takes on the value 0 and $W_{(n)}$ takes on the value 1 constantly.

We consider the maximum likelihood estimator of $\delta$ based on $W_{(1)}, W_{(2)}, \ldots, W_{(n)}$. The likelihood function of $\delta$ based on $W_{(1)}, W_{(2)}, \ldots, W_{(n)}$ might be bounded (will be proved later) since these are not dependent on $\gamma$ as mentioned above. To obtain the maximum likelihood estimator of $\delta$ based on $W_{(i)}$'s, we first derive the joint $PDF$ of $W_{(2)}, W_{(3)}, \ldots, W_{(n-1)}$.

**Proposition 1**. For $\delta > 0$, the joint density $PDF$ of $W_{(2)}, W_{(3)}, \ldots, W_{(n-1)}$ is given by

$$\psi(w_2, w, \ldots, w_{n-1}) = n! \delta^n \int_{-\infty}^{0} \int_{-\infty}^{0} (-u)^{n-2} \left\{ \prod_{i=1}^{n} (-v - u + uw_i) \right\}^{\delta-1}$$

$$\times exp\left[-\{\sum_{i=1}^{n}(-v - u + u\omega w_i)^\delta\}\right] du dv$$

when $0 \leq w_2 \leq \cdots \leq w_{n-1} \leq 1$; and $w_1 = 0$, $w_n = 1$.

**Proof**:

Denote $F(.;1,\delta,0)$ by $G(.;\delta)$ and $f(.;1,\delta,0)$ by $g(.;\delta)$, for simplicity. Suppose $Z_1, Z_2, \ldots, Z_n$ are $n$ independent random variables from the standard $RW$ distribution with $PDF$ $g(z_i;\delta)$ and $CDF$ $G(z_i;\delta)$. For $i = 1,2,\ldots,n$, let $Z_{(i)}$ be the i-th order statistics among $Z_1, Z_2, \ldots, Z_n$. For $n-2$ real value $0 \leq w_2 \leq \cdots \leq w_{n-1} \leq 1$, let us consider

$$P_r(W_{(i)} \leq w_i, i = 2,\ldots,n-1) = P_r\left(\frac{Z_{(i)} - Z_{(1)}}{Z_{(n)} - Z_{(1)}} \leq w_i, i = 2,\ldots,n-1\right)$$

$$= \int_{-\infty}^{0} \int_{-\infty}^{v} P_r(Z_{(i)} \leq u + (v-u)w_i, i = 2,\ldots,n-1 | Z_{(1)} = u, Z_{(n)} = v)$$

$$\times n(n-1)\, g(v;\delta)g(u;\delta)\{G(v;\delta) - G(u;\delta)\}^{n-2} du\, dv$$

$$= \int_{-\infty}^{0} \int_{-\infty}^{v} n!\, g(v;\delta)g(u;\delta) \prod_{i=2}^{n-1}\{G(u + (v-u)w_i;\delta)\} du\, dv. \qquad (6)$$

For every $u, v$ such that $u < v < 0$, $n > 2$ and $\delta > 0$, the integrand in (6), i.e,

$$n!\, g(v;\delta)g(u;\delta) \prod_{i=2}^{n-1}\{G(u + (v-u)w_i;\delta)\}$$

has a partial derivative

$$n!\, g(v;\delta)g(u;\delta) \prod_{i=2}^{n-1}(v-u)g(u + (v-u)w_i;\delta),$$

with respect to $w_i, i = 2,\ldots,n-1$. Moreover, we not that

$$\frac{(n-2)!\prod_{i=2}^{n-1}(v-u)g(u+(v-u)w_i;\delta)}{\{G(v;\delta) - G(u;\delta)\}^{n-2}},$$

is bounded above then we have

$$n!\prod_{i=2}^{n-1}(v-u)g(u+(v-u)w_i;\delta) \leq n(n-1)g(v;\delta)g(u;\delta)\{G(v;\delta)-G(u;\delta)\}^{n-2},$$

and by applying part(ii) of theorem 16.8 Billingsley[1], the partial derivative of

$$\text{Pr}(Wi \leq w_i; i = 2,\ldots,n-1)$$

with respect to $w_i\ i = 2,\ldots,n-1$, is given by

$$\int_{-\infty}^{0}\int_{-\infty}^{v} \frac{\partial^{n-2} n!\, g(v;\delta)g(u;\delta)\prod_{i=2}^{n-1} G(u+(v-u)w_i;\delta)}{\prod_{i=2}^{n-1}\partial w_i} dv\, du$$

$$= \int_{-\infty}^{0}\int_{-\infty}^{v} n!\, g(v;\delta)g(u;\delta)(v-u)^{n-2}\prod_{i=2}^{n-1} g(u+(v-u)w_i;\delta)\, dv\, du. \qquad (7)$$

For $w_2, \cdots, w_{n-1}$ for which $0 \leq w_2 \leq \cdots \leq w_{n-1} \leq 1$ is not satisfied, the partial derivative of $P_r(W_{(i)} \leq w_i, i = 2,..,n-1)$ with respect to $w_{(i)}, i = 2,....,n-1$, is always equal to 0 since

$$\lim_{\partial w_i \to 0, i=2,...,n-1} \frac{P_r(w_i \leq W_{(i)} \leq w_i + \delta w_i, i = 2,....,n-1)}{\prod_{i=2}^{n-1} \partial w_i} = 0.$$

After suitable transformations of variables $u$ and $v$, the proof of proposition 1 gets completed. □

From Proposition 1, we can obtain the likelihood function of $\delta$ based on $W_{(1)},....,W_{(n)}$ as

$$l(\delta; w2, w3, \ldots, wn-1) = \psi(w2, w3, \ldots, wn-1; \delta), \tag{8}$$

Where $w_2, w_3, ..., w_{n-1}$ are the realized values of $W_{(2)}, W_{(3)}, ..., W_{(n-1)}$. Then, the MLE of $\delta$ based on $W'_{(i)}s$, denoted by $\hat{\delta}_w$, is obtained by maximizing $l(\delta; w_2, w_3, \ldots, w_{n-1})$ with respect to $\delta$, by substituting $W'_{(i)}s$ for $w'_i s$.

**Proposition 2.** For $\delta > 0$ and any *given* $w_2, w_3, ..., w_{n-1}$ such that $0 \leq w_2 \leq \cdots \leq w_2 \leq 1$, the likelihood function $l(\delta; w_2, w_3, \ldots, w_{n-1})$ is differentiable with respect to $\delta$, and the derivative $l'(\delta; w_2, \ldots, w_{n-1})$ is given by

$$l'(\delta, w_2, \ldots, w_{n-1}) = n!\, \delta^n \int_{-\infty}^{0} \int_{-\infty}^{0} \left\{ \frac{n}{\delta} + \sum_{i=1}^{n} \log(-v - u + uw_i) [(-v - u + uw_i)^\delta] \right\} \exp\{(n-2)\log(-u) + (\delta - 1)\sum_{i=1}^{n} \log(-v - u + uw_i) - \sum_{i=1}^{n}(-v - u + uw_i)^\delta\} du\, dv.$$

**Proof:**

For $\delta > 0$, given $w_2, w_3, ..., w_{n-1}$ such that $0 \leq w_2 \leq \cdots \leq w_{n-1} \leq 1$, $l(\delta; w_2, w_3, \ldots, w_{n-1})$ can be rewritten as

$$l(\delta; w_2, \ldots, w_{n-1}) = n! \int_{-\infty}^{0} \int_{-\infty}^{0} \exp\{h(\delta, u, v; w_2, \ldots, w_{n-1})\} du\, dv,$$

where

$$h(\delta, u, v; w_2, \ldots, w_{n-1})$$
$$= n\log\delta + (n-2)\log(-u) + (\delta - 1)\sum_{i=1}^{n} \log(-v - u + uw_i) - \sum_{i=1}^{n}(-v - u + uw_i)^\delta$$

without loss of generality, we denote $h(\delta, u, v; w_2, \ldots, w_{n-1})$ by $h(\delta, u, v)$ in the remaining part of this proof. For every $\delta > 0, u < 0, v < 0, n > 2$ and $w_2, w_3, ..., w_{n-1}$ such that $0 \leq w_2 \leq \cdots \leq w_{n-1} \leq 1$, the partial derivative of $exp\{h(\delta, u, v)\}$ with respect to $\delta$ is given by $h'(\delta, u, v)exp\{h(\delta, u, v)\}$, where $h'(\delta, u, v)$, is the partial derivative of $h(\delta, u, v)$ with respect to $\delta$, which is

$$\frac{n}{\delta} + \sum_{i=1}^{n} \log(-v - u + uw_i)[1 - (-v - u + uw_i)^{\delta}].$$

And $|h'(\delta, u, v) exp\{h(\delta, u, v)\}|$ is bounded above and then there exists $M$ such that

$$|h'(\delta, u, v) exp\{h(\delta, u, v)/2\}| \leq M.$$

Thus

$$\int_{-\infty}^{0} \int_{-\infty}^{0} |h'(\delta, u, v) exp\{h(\delta, u, v)\}| \, dudv$$

$$\leq M \int_{-\infty}^{0} \int_{-\infty}^{0} exp\{h(\delta, u, v)/2\} \, dudv$$

$$\leq M \left( \int_{-\infty}^{0} \int_{-\infty}^{0} exp\{h(\delta, u, v)\} \, dudv \right)^{\frac{1}{2}}.$$

The second last inequality is due to Lyapunov's inequality while the last inequality holds from proposition 1.

Then, by applying part (ii) of theorem 16.8 of Billingsley [1], the derivative of $l(\delta; w_2, \ldots, w_{n-1})$ is given by

$$l'(\delta; w_2, \ldots, w_{n-1}) = n! \int_{-\infty}^{0} \int_{-\infty}^{0} \frac{\partial exp\{h(\delta, u, v)\}}{\partial \delta} dudv$$

$$= n! \int_{-\infty}^{0} \int_{-\infty}^{0} h'(\delta, u, v) exp\{h(\delta, u, v)\} \, dudv. \tag{9}$$

The proof of proposition 2 is thus complete. □

The following theorem and the using corollary implies that the estimate of $\delta$ obtained by maximizing $l(\delta; w_2, \ldots, w_{n-1})$ or solving equation $l'(\delta; w_2, \ldots, w_{n-1}) = 0$ always exists uniquely over the entire parameter space.

**Theorem' 1**. For $\delta > 0$ and any given $w_2, \ldots, w_{n-1}$ such that $0 \leq w_2 \leq \cdots \leq w_{n-1} \leq 1$ the likelihood equation $l'(\delta; w_2, \ldots, w_{n-1}) = 0$ always has a unique solution.

**Proof:**

First, we shall that likelihood equation has at least one solution. For simplicity, we denote $l'(\delta; w_2, \ldots, w_{n-1})$ by $l'(\delta)$, $h(\delta, u, v, w_2 \ldots, w_{n-1})$ by $h(\delta, u, v)$ and $h'(\delta, u, v, w_2 \ldots, w_{n-1})$ by $h'(\delta, u, v)$ in the remaining part of this proof. Since $exp\{h(\delta, u, v)\} > 0$, $\lim_{\delta \downarrow 0} h'(\delta, u, v) = \infty$, and $\lim_{\delta \to \infty} h'(\delta, u, v) < 0$ for every $\delta > 0$ and $u < 0$ and $v < 0$, there exist a positive real value $\delta_1$, such that $l'(\delta) > 0$ for all $\delta$ in $(0, \delta1)$ and a positive real value $\delta_2$, such that $l'(\delta) < 0$ for all $\delta > \delta_2$, also for $\delta > 0$, $l'(\delta)$ is continuous with respect to $\delta$. Thus $l'(\delta) = 0$ always has at least one solution. Next we shall show that the number of solution exactly one. Let, $s_{u,v,i:n} = \log(-u - v + uw_i), i = 1, \ldots, n$ for simplicity. Then, we rewrite $h(\delta, u, v)$ and $h'(\delta, u, v)$ as

$$h(\delta, u, v) = h(\delta, S_{u,v}) = n\log\delta + (n-2)\log(-u) + (\delta - 1)\sum_{i=1}^{n} s_{u,v,i:n} - \sum_{i=1}^{n} \exp(\delta s_{u,v,i:n})$$

and

$$h'(\delta, u, v) = h'(\delta, S_{u,v}) = h'\left(\delta, \xi(\beta, S_{u,v})\right)$$

$$= \frac{n}{\delta} + \sum_{i=1}^{n} s_{u,v,i:n} - \sum_{i=1}^{n} s_{u,v,i:n} \exp(\delta s_{u,v,i:n})$$

$$= \frac{n}{\delta} + \sum_{i=1}^{n} \{1 - \exp(\delta s_{u,v,i:n})\} s_{u,v,i:n}$$

$$= \frac{n}{\delta} + \xi(\delta, S_{u,v}),$$

where $S_{u,v} = (s_{u,v,1:n}, s_{u,v,2:n}, \ldots, s_{u,v,n:n})$ and

$$\xi(\delta, S_{u,v}) = \sum_{i=1}^{n} \{1 - \exp(\delta) s_{u,v,i:n}\} s_{u,v,i:n}, \qquad i = 1, 2, \ldots, n.$$

we see that each $s_{u,v,i:n}, i = 1, 2, \ldots, n$, takes on values over $(-\infty, +\infty)$ for $u$ and $v<0$, note that, if $s_{(u,v,i:n)} < 0, i = 1, \ldots, n$, $\xi(\delta, Su, v)$ is strictly increasing in each $S_{u,v,i:n} < 0, i = 1, \ldots, n$, and takes on value over $(-\infty, 0)$, for any fixed $\delta > 0$, thus there exist a unique value of $\xi(\delta, S_{u,v})$ on the set $\{S_{u,v}: s_{u,v,i:n} < 0, i = 1, \ldots, n\}$ such that $h'(\delta, S_{u,v}) = 0$, for any fixed $\delta > 0$, we denote the value of $\xi(\delta, S_{u,v})$ by $\xi_0^-(\delta)$, we see that $h'(\delta, S_{u,v}) < 0$ for $S_{u,v}$ on the set $\{Su, v; \xi(\delta, S_{u,v}) < \xi_0^-(\delta), S_{u,v}, i: n < 0, i = 1, \ldots, n\}$ and for $h'(\delta, S_{u,v}) > 0$ for $S_{u,v}$ on the set $\{S_{u,v}; \xi(\delta, S_{u,v}) > \xi_0^-(\delta), S_{u,v,i:n} < 0, i = 1, \ldots, n\}$ for any $\delta > 0$. Analogously, if $S_{u,v,i:n} > 0, i = 1, \ldots, n; \xi(\delta, S_{u,v})$ is strictly decreasing in each $S_{u,v,i:n} < 0; i = 1, \ldots, n$ and take on values over $(-\infty, 0)$ thus, there exists a unique value of $\xi(\delta, S_{u,v})$ on the set $\{S_{u,v}: s_{u,v,i:n} > 0; i = 1, \ldots, n\}$ such that $h'(\delta, S_{u,v}) = 0$, for any fixed $\delta > 0$.we denote the value of $\xi(\delta, S_{u,v})$ by $\xi_0^+(\delta)$. We see that $h'(\delta, S_{u,v}) > 0$ for $S_{u,v}$ on the set $\{S_{u,v}; \xi(\delta, S_{u,v}) > \xi_0^+(\delta), S_{u,v,i:n} > 0, i = 1, \ldots, n\}$ and that $h'(\delta, S_{u,v}) < 0$ for $S_{u,v}$ on the set $\{S_{u,v}; \xi(\delta, S_{u,v}) < \xi_0^+(\delta), S_{u,v,i:n} > 0, i = 1, \ldots, n\}$ for any $\delta > 0$.
Let, for $\Delta\delta$,

$$\Phi(\delta, \Delta\delta, S_{u,v}) = \frac{h'(\delta + \Delta\delta, \xi(\delta, S_{u,v})) \exp\{h(\delta + \Delta\delta, S_{u,v})\}}{h'(\delta, \xi(\delta, S_{u,v})) \exp\{h(\delta, S_{u,v})\}}$$

$$= \frac{\frac{n}{\delta + \Delta\delta} + \xi(\delta + \Delta\delta, S_{u,v})}{\frac{n}{\delta} + \xi(\delta, S_{u,v})} (1 + \frac{\Delta\delta}{\delta})^n$$

$$\times exp\left\{\Delta\delta\sum_{i=1}^{n} s_{u,v,i:n} + \sum_{i=1}^{n}[1-\exp(\Delta\delta s_{u,v,i:n})]\exp(\delta s_{u,v,i:n})\right\}; \quad (10)$$

Then $l'(\delta + \Delta\delta)$ can be rewritten as

$$l'(\delta + \Delta\delta) = n! \int_{-\infty}^{0}\int_{-\infty}^{0} \Phi(\delta,\Delta\delta,S_{u,v}) h'(\delta,S_{u,v}) \exp\{h(\delta,S_{u,v})\} du dv. \quad (11)$$

From now on, let us focus on the case when $\Delta\delta \geq 0$. We note that,

$$\lim_{\Delta\delta \downarrow 0} \Phi(\delta,\Delta\delta,S_{u,v}) = \Phi(\delta,0,S_{u,v}) = 1,$$

For

$$(u,v)\in\{\xi(\delta,S_{u,v}) \neq \xi_0^-(\delta) \text{ and } \xi_0^+(\delta)\}. \quad (12)$$

While

$$\Phi(\delta,\Delta\delta,S_{u,v}) = \begin{cases} 1 & \text{if } \Delta\delta = 0 \\ -\infty & \text{if } \Delta\delta > 0 \end{cases} \quad (u,v)\in\{\xi(\delta,S_{u,v}) = \xi_0^-(\delta) \text{ or } \xi_0^+(\delta)\}. \quad (13)$$

for any $\delta$.

Let $\delta^*$ be one of the solutions of $l'(\delta) = 0$. Then,

$$\lim_{\Delta\delta \downarrow 0}\frac{l'(\delta^* + \Delta\delta) - l'(\delta^*)}{\Delta\delta} = \lim_{\Delta\delta \downarrow 0}\frac{l'(\delta^* + \Delta\delta)}{\Delta\delta}$$

$$= \lim_{\Delta\delta \downarrow 0}\frac{n! \iint_{\{u<0,v<0\}} \Phi(\delta^*,\Delta\delta,S_{u,v}) h'(\delta^*,S_{u,v})\exp\{h(\delta^*,S_{u,v})\} du dv}{\Delta\delta}$$

$$= n! \lim_{\Delta\delta \downarrow 0}\frac{1}{\Delta\delta}[\iint_{\{\xi(\delta^*,S_{u,v})=\xi_0^-(\delta^*) \text{ or } \xi_0^+(\delta^*)\}} \Phi(\delta^*,\Delta\delta,S_{u,v}) h'(\delta^*,S_{u,v})\exp\{h(\delta^*,S_{u,v})\} du dv$$

$$+ \iint_{\{\xi(\delta^*,S_{u,v}) \neq \xi_0^-(\delta^*) \text{ and } \xi_0^+(\delta^*)\}} \Phi(\delta^*,\Delta\delta,S_{u,v}) h'(\delta^*,S_{u,v})\exp\{h(\delta^*,S_{u,v})\} du dv]. \quad (14)$$

From (12) and (13) it is too easy to see that $\Phi(\delta^*,\Delta\delta,S_{u,v})$, for any $(u,v)\in\{\xi(\delta^*,S_{u,v}) \neq \xi_0^-(\delta^*) \text{ and } \xi_0^+(\delta^*)\}$ approaches 1 faster than $\Phi(\delta^*,\Delta\delta,S_{u,v})$ for any $(u,v)\in\{\xi(\delta^*,S_{u,v}) = \xi_0^-(\delta^*) \text{ or } \xi_0^+(\delta^*)\}$ approaching 1 when $\Delta\delta$ decreases. Hence

$$\iint_{\{\xi(\delta^*,S_{u,v}) \neq \xi_0^-(\delta^*) \text{ and } \xi_0^+(\delta^*)\}} \Phi(\delta^*,\Delta\delta,S_{u,v}) h'(\delta^*,S_{u,v})\exp\{h(\delta^*,S_{u,v})\} du dv$$

approaches

$$\iint_{\{\xi(\delta^*,S_{u,v}) \neq \xi_0^-(\delta^*) \text{ and } \xi_0^+(\delta^*)\}} h'(\delta^*,S_{u,v})\exp\{h(\delta^*,S_{u,v})\} du dv$$

faster that

$$\iint_{\{\xi(\delta^*,S_{u,v})=\xi_0^-(\delta^*) \text{ or } \xi_0^+(\delta^*)\}} \Phi(\delta^*,\Delta\delta,S_{u,v})h'(\delta^*,S_{u,v})exp\{h(\delta^*,S_{u,v})\}dudv$$

approaching

$$\iint_{\{\xi(\delta^*,S_{u,v})=\xi_0^-(\delta^*) \text{ or } \xi_0^+(\delta^*)\}} h'(\delta^*,S_{u,v})exp\{h(\delta^*,S_{u,v})\}dudv,$$

when $\Delta\delta$ decreases. Note further than

$$\iint_{\{\xi(\delta^*,S_{u,v})\neq\xi_0^-(\delta^*) \text{ and } \xi_0^+(\delta^*)\}} h'(\delta^*,S_{u,v})exp\{h(\delta^*,S_{u,v})\}dudv$$

$$= \iint_{\{\xi(\delta^*,S_{u,v})=\xi_0^-(\delta^*) \text{ or } \xi_0^+(\delta^*)\}} h'(\delta^*,S_{u,v})exp\{h(\delta^*,S_{u,v})\}dudv = 0$$

Therefore by the fundamental theory of differential calculus, the sign of the RHS of (14) agrees with sign of

$$\iint_{\{\xi(\delta^*,S_{u,v})=\xi_0^-(\delta^*) \text{ or } \xi_0^+(\delta^*)\}} \Phi(\delta^*,\Delta\delta,S_{u,v})h'(\delta^*,S_{u,v})exp\{h(\delta^*,S_{u,v})\}dudv$$

for sufficiently small $\Delta\delta > 0$, which implies

$$\lim_{\Delta\delta\downarrow 0} \frac{l'(\delta^* + \Delta\delta) - l'(\delta^*)}{\Delta\delta} < 0 \qquad (15)$$

Since for any

$$\Delta\delta > 0, h'(\delta^* + \Delta\delta, S_{u,v}) < 0$$

And

$$exp\{\xi(\delta^* + \Delta\delta, S_{u,v})\} > 0$$

for any

$$(u,v)\epsilon\{\xi(\delta^*,S_{u,v}) = \xi_0^-(\delta^*) \text{ or } \xi_0^+(\delta^*)\}$$

and thus

$$\iint_{\{\xi(\delta^*,S_{u,v})=\xi_0^-(\delta^*) \text{ or } \xi_0^+(\delta^*)\}} \Phi(\delta^*,\Delta\delta,S_{u,v})h'(\delta^*,S_{u,v})exp\{h(\delta^*,S_{u,v})\}dudv$$

$$= \iint_{\{\xi(\delta^*,S_{u,v})=\xi_0^-(\delta^*) \text{ or } \xi_0^+(\delta^*)\}} h'(\delta^* + \Delta\delta, S_{u,v})exp\{h(\delta^* + \Delta\delta, S_{u,v})\}dudv < 0$$

Analogously, we obtain the following inequality:

$$\lim_{\Delta\delta\uparrow 0} \frac{l'(\delta^* + \Delta\delta) - l'(\delta^*)}{\Delta\delta} < 0 \qquad (16)$$

The proof is very similar to proof of (15) and is therefore omitted for the sake of brevity.

It follows from (15), (16) and the fact that $l'(\delta)$ is differentiable with respect to $\delta$ (the proof is very similar to the proof of the differentiability of $l(\delta;w_2,...,w_{n-1})$ in proposition 2 and is therefore omitted for the sake of brevity that $\frac{l'(\delta^*)}{\Delta\delta} < 0$ holds.

This is clearly implies that $l'(\delta)$ changes sign only once with respect to $\delta$. From the facts established above, $l'(\delta) = 0$ always has a unique solution with respect to $\delta$. The proof of theorem 1 is thus completed. □

**Corollary 1.** For $\delta > 0$, and any given $w_2, w_3, \ldots, w_{n-1}$ such that $0 \leq w_2 \leq \cdots \leq w_{n-1} \leq 1$, the likelihood function $l(\delta; w_2, w_3, \ldots, w_{n-1})$ is unimodal with respect to $\delta$.

**Proof**: Corollary 1 the obvious from theorem 1, and the proof therefore omitted. □

One of the main purposes of this section is to prove that the estimate of $\delta$ has consistency. The following lemma is needed before presenting the result about the consistency.

**Lemma 1.** For any fixed $\delta \neq \delta_0$, where $\delta_0$ is the true value of the parameter $\delta$,

$$\lim_{n \to \infty} P_r(l(\delta; W_{(2)}, \ldots W_{(n)}) < l(\delta_0; W_{(2)}, \ldots, W_{(n)})) = 1$$

**Proof**:

Let $W_i, i = 2, \ldots, n-1$, be the random variables whose order statistics are $W_{(i)}, i = 2, \ldots, n-1$. For every $u$ and $v$ such that $u < v < 0$, under the conditions $Z_{(1)} = u, Z_{(n)} = v$, where $Z_{(1)} = \frac{X_{(1)} - \gamma_0}{\beta_0}, Z_{(n)} = \frac{X_{(n)} - \gamma_0}{\beta_0}$, and $\beta_0$ and $\gamma_0$ are the true values of $\beta$ and $\gamma$, respectively, since the conditional joint pdf of the order statisticts of $W_i$'s given $Z_{(1)} = u, Z_{(n)} = v$, is given by

$$(n-2)! \prod_{i=2}^{n-1}(v-u) g(u + (v-u)w_i; \delta_0)/\{G(v; \delta_0) - G(u; \delta_0)\}, \quad 0 \leq w_2 \leq \cdots \leq w_{n-1} \leq 1$$

$W_i, i = 2, \ldots, n-1$, are distributed with the common conditional pdf, given $Z_{(1)} = u, Z_{(n)} = v$, which is expressed as

$$(v-u)g(u + (v-u)w_i; \delta_0)/\{G(v; \delta_0) - G(u; \delta_0)\}, \quad 0 \leq w_i \leq 1. \tag{17}$$

and these are conditionally independent, given $Z_{(1)} = u, Z_{(n)} = v$.

Denote

$$(n-2)! \prod_{i=2}^{n-1}(v-u) g(u + (v-u)w_i; \delta)/\{G(v; \delta) - G(u; \delta)\}$$

by $l_{u,v}(\delta; W_{(2)}, \ldots, W_{(n-1)})$. For any fixed $u$ and $v$ such that $u < v < 0$, under the conditional $Z_{(1)} = u, Z_{(n)} = v$, let us consider, for every $\delta \neq \delta_0, n > 2$,

$$\frac{1}{n-2}\left[\log \frac{l_{u,v}(\delta; W_{(2)}, \ldots, W_{(n-1)})}{l_{u,v}(\delta_0; W_{(2)}, \ldots, W_{(n-1)})}\right]$$

$$= \frac{1}{n-2} \sum_{i=2}^{n-1} \log \left[\frac{(v-u)g(u + (v-u)w_i; \delta)/\{G(v; \delta) - G(u; \delta)\}}{(v-u)g(u + (v-u)w_i; \delta_0)/\{G(v; \delta_0) - G(u; \delta_0)\}}\right] \tag{18}$$

By the law of large numbers, (18) tends in probability to

$$E\left[\log \left[\frac{(v-u)g(u + (v-u)w; \delta)/\{G(v; \delta) - G(u; \delta)\}}{(v-u)g(u + (v-u)w; \delta_0)/\{G(v; \delta_0) - G(u; \delta_0)\}}\right]\right] \tag{19}$$

Where $W$ is a random variable which is distributed with the conditional $PDF$ in (17), given $Z_{(1)} = u$, $Z_{(n)} = v$. By Jensen's inequality, we have

$$E\left[\log\left[\frac{(v-u)g(u+(v-u)w;\delta)/\{G(v;\delta)-G(u;\delta)\}}{(v-u)g(u+(v-u)w;\delta_0)/\{G(v;\delta_0)-G(u;\delta_0)\}}\right]\right]$$

$$< \log\left\{E\left[\frac{(v-u)g(u+(v-u)w;\delta)/\{G(v;\delta)-G(u;\delta)\}}{(v-u)g(u+(v-u)w;\delta_0)/\{G(v;\delta_0)-G(u;\delta_0)\}}\right]\right\}$$

$$= \log\int_0^1 \frac{(v-u)g(u+(v-u)w;\delta)}{\{G(v;\delta)-G(u;\delta)\}} = 0 \qquad (20)$$

It follows that

$$\lim_{n\to\infty} P_r\left\{\frac{1}{n-2}\log\frac{l_{u,v}(\delta;W_{(2)},\dots,W_{(n-1)})}{l_{u,v}(\delta_0;W_{(2)},\dots,W_{(n-1)})} < 0 \middle| Z_{(1)}=u, Z_{(n)}=v\right\} = 1$$

or

$$\lim_{n\to\infty} P_r\{l_{u,v}(\delta;W_{(2)},\dots,W_{(n-1)}) < l_{u,v}(\delta_0;W_{(2)},\dots,W_{(n-1)})|Z_{(1)}=u, Z_{(n)}=v\} = 1 \qquad (21)$$

By the positivity and the integrability of $l_{u,v}(\delta;W_{(2)},\dots,W_{(n-1)})$ and $l_{u,v}(\delta_0;W_{(2)},\dots,W_{(n-1)})$ and (21), implies

$$\lim_{n\to\infty} P_r\{l(\delta;W_{(2)},\dots,W_{(n-1)}) < l(\delta_0;W_{(2)},\dots,W_{(n-1)})|Z_{(1)}=u, Z_{(n)}=v\} = 1 \qquad (22)$$

Moreover

$$\int_{-\infty}^0\int_{-\infty}^v n(n-1)g(u;\delta_0)g(v;\delta_0)\{G(v;\delta_0)-G(u;\delta_0)\}^{n-2}\,dudv = 1 \qquad (23)$$

and

$$\int_{-\infty}^0\int_{-\infty}^v n(n-1)g(u;\delta_0)g(v;\delta_0)\{G(v;\delta_0)-G(u;\delta_0)\}^{n-2}$$

$$\times P_r\{l(\delta;W_{(2)},\dots W_{(n)}) < l(\delta_0;W_{(2)},\dots,W_{(n)})|Z_{(1)}=u, Z_{(n)}=v\}dudv$$

$$\leq \int_{-\infty}^0\int_{-\infty}^v n(n-1)g(u;\delta_0)g(v;\delta_0)\{G(v;\delta_0)-G(u;\delta_0)\}^{n-2}\,dudv = 1 \qquad (24)$$

since $P_r\{l(\delta;W_{(2)},\dots W_{(n)}) < l(\delta_0;W_{(2)},\dots,W_{(n)})|Z_{(1)}=u, Z_{(n)}=v\}$ is bounded by 1.

Then by applying the dominated convergence theorem, from (23) and (24) it follows that

$$\lim_{n\to\infty} P_r\left(l(\delta;W_{(2)},\dots,W_{(n-1)}) < l(\delta_0;W_{(2)},\dots,W_{(n-1)})\right)$$

$$= \int_{-\infty}^{0} \int_{-\infty}^{v} \lim_{n \to \infty} n(n-1)g(u; \delta_0)g(v; \delta_0)\{G(v; \delta_0) - G(u; \delta_0)\}^{n-2}$$

$$\times P_r\{l(\delta; W_{(2)}, \ldots W_{(n)}) < l(\delta_0; W_{(2)}, \ldots, W_{(n)}) | Z_{(1)} = u, Z_{(n)} = v\} du dv$$

$$= \int_{-\infty}^{0} \int_{-\infty}^{v} \lim_{n \to \infty} n(n-1)g(u; \delta_0)g(v; \delta_0)\{G(v; \delta_0) - G(u; \delta_0)\}^{n-2} du dv$$

$$= \lim_{n \to \infty} \int_{-\infty}^{0} \int_{-\infty}^{v} n(n-1)g(u; \delta_0)g(v; \delta_0)\{G(v; \delta_0) - G(u; \delta_0)\}^{n-2} du dv = 1$$

The proof of Lemma 1 is thus complete. □

**Theorem 2.** The estimator $\hat{\delta}_w$ is consistent for $\delta > 0$.

**Proof:** The proof is very similar to the proof of Theorem 3.7 of Lehmann and Casella [11], and is therefore omitted. □

### 3.3. Estimation of location and scale parameters

Once we obtain the estimate of $\delta$, using the method outlined above, we may proceed to the stimation of $\gamma$ and $\beta$, where in the estimators have the following properties.

**Property 1.** The estimates exist uniquely for all n and for all $\delta, \gamma$ and $\beta$, where $n > 2, \delta > 0, -\infty < \gamma < \infty$ and $\beta > 0$.

**Property 2.** The estimators are consistent for $\gamma$ and $\beta$, respectively.

First, before providing the estimators having the above properties, we consider the following estimators of $\gamma$ and $\beta$:

$$\hat{\gamma}_{init} = X_{(n)} \tag{25}$$

and

$$\hat{\beta}_{init} = \left[\frac{\sum_{i=1}^{n}(\hat{\gamma}_{init} - x_i)^{\hat{\delta}_W}}{n}\right]^{\frac{1}{\hat{\delta}_W}} \tag{26}$$

It is evident that the estimates $\hat{\gamma}_{init}$ and $\hat{\beta}_{init}$ uniquely exist, given the observations $x_1, \ldots, x_n$, where $\hat{\delta}'_W$ is the realized value of $\hat{\delta}_W$. It is well-known that $X_{(n)}$ tends in probability to $\gamma$ as $n \to \infty$ for every $\gamma$ since

$$E(X_{(n)} - \gamma)^2 = \frac{n\delta}{\beta}\int_{-\infty}^{\gamma}(\gamma - x)^2 \exp\left[-n\left(\frac{\gamma - x}{\beta}\right)^{\delta}\right]\left(\frac{\gamma - x}{\beta}\right)^{\delta-1} dx$$

when $n\left(\frac{\gamma-x}{\beta}\right)^{\delta} = z$ then $E(X_{(n)} - \gamma)^2 = \beta^2 \int_0^\infty \sqrt[\delta]{\frac{z^2}{n^2}} e^{-z} dz$, when $\to \infty$; $E(X_{(n)} - \gamma)^2 = 0$.

Assuming that $\delta$ and $\gamma$ are known and substituting $\delta$ for $\hat{\delta}_W$ and $\gamma$ for $\hat{\gamma}_{init}$ in (25,26), $\hat{\beta}_{init}$ is the maximum likelihood estimator of $\beta$ in the regular case and therefore consistent for $\beta$. It follows these facts and slutsky´s theorem that $\hat{\beta}_{init}$ is consistent for $\beta$. The estimators $\hat{\gamma}_{init}$ and $\hat{\beta}_{init}$ then have properties 1 and 2 mentioned above. However, the estimators could have considerable bias since $\hat{\gamma}_{init}$ has significant bias. So, we need to consider correction of bias for these estimators.

Since

$$E[X_{(n)}] = \gamma - \beta\left(1 + \frac{1}{\delta}\right)n^{-\frac{1}{\delta}}$$

It is easy to proof, upon substituting $\hat{\delta}_W$ for $\delta$ and $\hat{\beta}_{init}$ for $\beta$, the bias-corrected estimator of $\gamma$ becomes

$$\hat{\gamma}_W = X_{(n)} + \hat{\beta}_{init}\left(1 + \frac{1}{\hat{\delta}_W}\right)n^{-\frac{1}{\hat{\delta}_W}}. \tag{27}$$

We then obtain the bias-corrected estimator of $\beta$ as

$$\hat{\beta}_W = \left[\frac{\sum_{i=1}^{n}(\hat{\gamma}_W - x_i)^{\hat{\delta}_W}}{n}\right]^{\frac{1}{\hat{\delta}_W}}$$

It follows, from the above mentioned forms and the fact that the term $\hat{\beta}_{init}\left(1 + \frac{1}{\hat{\delta}_W}\right)n^{-\frac{1}{\hat{\delta}_W}}$ in (27) tends in probability to 0 as $n \to \infty$ (which can be shown easily by using slutsky´s theorem), that $\hat{\gamma}_W$ and $\hat{\beta}_W$ also have properties 1 and 2.

### 4. Simulation

We carry out a Mont Carlo simulation study to evaluate the compare of the proposed estimators. The proposed estimators, termed LSPFE, MLE and MME.

In the simulation study, the values of the shape parameter $\delta$ are selected as 0.5,1,2,3,4,5 and take $\gamma = 0$ and $\beta = 10$ the sample size is taken to be 20,50,100.

All programs in this numerical study were written in the package R. Tables 1-6 display the simulation results of the bias and root mean squared on 100 Monte carlo runs for each set of configurations. Bias column is joint columns represent the sum of the absolute values of bias of the three estimators, and $RMSE$ column in joint columns represents the root of the trace of $MSE$ matrix of the three estimators, which are used for evaluting the marginal performance based on bias and $RMSE$ of the three estimators. Figures 2-10 show the bias and $MRSE$ of tables 1-6. from these results, we observe the following.

**Table 1.** $Bias$, $RMSE$ of $LSPF$, $ML$, $MM$ estimators based on 100 simulations with $\delta = 0.5$ and $n = 20, 50, 100$.

| $\delta$ | n | method | location | | shape | | scale | | joint | |
|---|---|---|---|---|---|---|---|---|---|---|
| | | | Bias | RMSE | Bias | RMSE | Bias | RMSE | Bias | RMSE |
| 0.5 | 20 | LSPF | -0.1610 | 0.5483 | -0.0431 | 0.7763 | 0.2325 | 0.5881 | 0.0095 | 1.1177 |
| | | MLE | -0.5512 | 0.5463 | 0.0091 | 0.5123 | -0.204 | 0.7746 | -0.2481 | 1.0775 |
| | | MME | -0.5971 | 0.7107 | -0.1231 | 0.7617 | 1.4276 | 2.3191 | 0.2358 | 2.5423 |
| | 50 | LSPF | 0.1375 | 0.5235 | -0.0471 | 0.6604 | 0.1664 | 0.4469 | 0.0856 | 0.9539 |
| | | MLE | 0.2975 | 0.5380 | 0.0231 | 0.4406 | 0.1959 | 0.6311 | 0.1721 | 0.9391 |
| | | MME | -0.5525 | 0.6965 | -0.1161 | 0.9375 | 1.1753 | 1.8561 | 0.1689 | 2.1931 |
| | 100 | LSPF | -0.1166 | 0.5051 | -0.0451 | 0.7485 | -0.0011 | 0.0008 | -0.0541 | 0.9030 |
| | | MLE | -0.1833 | 0.5076 | -0.0661 | 0.6128 | -0.0011 | 0.0011 | -0.0831 | 0.7957 |
| | | MME | -0.5466 | 0.5507 | -0.0841 | 0.4406 | 1.2963 | 1.8461 | 0.2219 | 1.9762 |

**Table 2.** $Bias$, $RMSE$ of $LSPF$, $ML$, $MM$ estimators based on 100 simulations with $\delta = 1$ and $n = 20, 50, 100$.

| $\delta$ | n | method | location | | shape | | scale | | joint | |
|---|---|---|---|---|---|---|---|---|---|---|
| | | | Bias | RMSE | Bias | RMSE | Bias | RMSE | Bias | RMSE |
| 1 | 20 | LSPF | 0.3071 | 0.4804 | -0.0751 | 0.3331 | 0.4372 | 0.7175 | 0.2230 | 0.9255 |
| | | MLE | -0.2617 | 0.8345 | -0.2693 | 0.5987 | -0.4641 | 0.7351 | -0.3303 | 1.2631 |
| | | MME | -0.8715 | 1.0553 | -0.6748 | 1.0009 | 1.6422 | 2.1225 | 0.0318 | 2.5731 |
| | 50 | LSPF | 0.3005 | 0.4769 | -0.2021 | 0.5098 | -0.0131 | 0.0551 | 0.0295 | 0.7003 |
| | | MLE | 0.2387 | 0.7235 | -0.4966 | 0.7139 | -0.0621 | 0.1121 | -0.1059 | 1.0227 |
| | | MME | -0.8476 | 0.8930 | -0.6274 | 1.0793 | 1.1431 | 1.9301 | -0.1107 | 2.3849 |
| | 100 | LSPF | -0.1676 | 0.1974 | 0.0996 | 0.7273 | 0.0082 | 0.0223 | -0.0199 | 0.7541 |
| | | MLE | -0.2291 | 0.5246 | -0.2066 | 0.8295 | -0.0112 | 0.0265 | -0.1485 | 0.9818 |
| | | MME | -0.8158 | 0.8786 | -0.7266 | 0.9551 | 1.1161 | 1.3683 | -0.1422 | 1.8859 |

**Table 3.** $Bias$, $RMSE$ of $LSPF$, $ML$, $MM$ estimators based on 100 simulations with $\delta = 2$ and $n = 20, 50, 100$.

| $\delta$ | n | method | location | | shape | | scale | | joint | |
|---|---|---|---|---|---|---|---|---|---|---|
| | | | Bias | RMSE | Bias | RMSE | Bias | RMSE | Bias | RMSE |
| 2 | 20 | LSPF | -0.2212 | 0.9367 | -0.4791 | 1.0354 | -0.3491 | 0.3505 | -0.3496 | 1.4396 |
| | | MLE | -0.3448 | 0.8677 | -0.9122 | 0.9444 | -0.4692 | 0.4712 | -0.5752 | 1.3663 |
| | | MME | -0.9233 | 1.0066 | -1.7770 | 1.9459 | 1.1964 | 2.8355 | -0.5013 | 3.5833 |
| | 50 | LSPF | 0.1775 | 0.7071 | -0.4071 | 0.8143 | -0.2104 | 0.2523 | -0.1465 | 1.1076 |
| | | MLE | -0.2956 | 0.6685 | -0.7533 | 1.1845 | -0.2875 | 0.3550 | -0.4452 | 1.4057 |
| | | MME | -0.8775 | 0.9784 | -1.2994 | 1.9909 | 1.1918 | 1.7498 | -0.3282 | 2.8254 |
| | 100 | LSPF | 0.1334 | 0.1201 | -0.1352 | 0.4837 | -0.0624 | 0.1035 | -0.0211 | 0.5091 |
| | | MLE | 0.1066 | 0.5962 | -0.9841 | 1.3548 | -0.1422 | 0.1593 | -0.3397 | 1.4887 |
| | | MME | -0.7023 | 0.9346 | -1.4392 | 1.9803 | 1.1033 | 1.5260 | -0.3460 | 2.6691 |

**Table 4.** *Bias*, *RMSE* of *LSPF*, *ML*, *MM* estimators based on 100 simulations with $\delta = 3$ and $n = 20, 50, 100$.

| $\delta$ | n | method | Location | | shape | | scale | | joint | |
|---|---|---|---|---|---|---|---|---|---|---|
| | | | Bias | RMSE | Bias | RMSE | Bias | RMSE | Bias | RMSE |
| 3 | 20 | LSPF | -0.4035 | 0.4207 | -1.3051 | 2.3454 | -0.4162 | 0.4272 | -0.7081 | 2.4208 |
| | | MLE | -0.4816 | 0.8843 | -1.2705 | 2.4899 | -0.5883 | 0.5913 | -0.7798 | 2.7076 |
| | | MME | -0.9766 | 1.0329 | -2.1584 | 3.0973 | 1.58333 | 2.5923 | -0.5171 | 4.1689 |
| | 50 | LSPF | -0.1975 | 0.4092 | -1.1972 | 1.3112 | -0.2362 | 0.3702 | -0.5435 | 1.4226 |
| | | MLE | -0.3425 | 0.5783 | -1.4134 | 1.5540 | -0.2221 | 0.4404 | -0.6591 | 1.7156 |
| | | MME | -0.8965 | 0.9265 | -2.2703 | 2.8085 | 1.42134 | 1.8190 | -0.5817 | 3.4720 |
| | 100 | LSPF | 0.1885 | 0.2859 | -0.5832 | 1.1419 | -0.1593 | 0.3682 | -0.1844 | 1.2334 |
| | | MLE | -0.2633 | 0.3314 | -1.8962 | 2.0324 | -0.1844 | 0.4222 | -0.7811 | 2.1021 |
| | | MME | -0.7198 | 0.9956 | -2.2501 | 2.7733 | 1.41525 | 1.6870 | -0.5182 | 3.3954 |

**Table 5.** *Bias*, *RMSE* of *LSPF*, *ML*, *MM* estimators based on 100 simulations with $\delta = 4$ and $n = 20, 50, 100$.

| $\delta$ | n | method | locatin | | shape | | scale | | joint | |
|---|---|---|---|---|---|---|---|---|---|---|
| | | | Bias | RMSE | Bias | RMSE | Bias | RMSE | Bias | RMSE |
| 4 | 20 | LSPF | -0.5152 | 0.5994 | -1.7012 | 2.9138 | -0.4666 | 0.5766 | -0.8941 | 3.0301 |
| | | MLE | -0.5823 | 0.6881 | -2.1624 | 3.0818 | -0.3565 | 0.6814 | -1.0334 | 3.2303 |
| | | MME | -0.9361 | 0.9962 | -2.8055 | 4.2989 | 1.8166 | 2.8170 | -0.6412 | 5.2353 |
| | 50 | LSPF | -0.4353 | 0.4948 | -1.5972 | 2.3024 | -0.3864 | 0.5237 | -0.8062 | 2.4125 |
| | | MLE | -0.5021 | 0.5903 | -2.3731 | 2.4607 | -0.3365 | 0.6073 | -1.0703 | 2.6023 |
| | | MME | -0.7433 | 0.8967 | -3.1522 | 3.7828 | 1.4953 | 2.0957 | -0.7992 | 4.4165 |
| | 100 | LSPF | -0.3956 | 0.4070 | -0.7283 | 1.4701 | -0.2830 | 0.4881 | -0.4692 | 1.6015 |
| | | MLE | -0.4732 | 0.4163 | -2.0125 | 2.6382 | 0.2865 | 0.5355 | -0.7321 | 2.7240 |
| | | MME | -0.6324 | 0.7384 | -2.9104 | 3.7994 | 1.0646 | 1.8203 | -0.8254 | 4.2771 |

**Table 6.** *Bias*, *RMSE* of *LSPF*, *ML*, *MM* estimators based on 100 simulations with $\delta = 5$ and $n = 20, 50, 100$.

| $\delta$ | n | method | Location | | shape | | scale | | joint | |
|---|---|---|---|---|---|---|---|---|---|---|
| | | | Bias | RMSE | Bias | RMSE | Bias | RMSE | Bias | RMSE |
| 5 | 20 | LSPF | -0.6162 | 0.8932 | -2.4394 | 3.5809 | -0.4065 | 0.6076 | -1.1532 | 3.7403 |
| | | MLE | -0.6511 | 0.9086 | -2.6842 | 4.2845 | -0.5026 | 0.6279 | -1.2794 | 4.4246 |
| | | MME | -0.9133 | 1.1093 | -3.3283 | 4.9113 | 1.5332 | 3.4707 | -0.9023 | 6.1153 |
| | 50 | LSPF | -0.4525 | 0.4521 | -2.2532 | 3.1435 | -0.3915 | 0.5493 | -1.0322 | 3.2230 |
| | | MLE | -0.5271 | 0.5284 | -2.5521 | 3.8372 | -0.4294 | 0.5873 | -1.1693 | 3.9176 |
| | | MME | -0.7895 | 0.9893 | -3.4084 | 4.8109 | 1.4839 | 2.4489 | -0.9045 | 5.4882 |
| | 100 | LSPF | -0.4502 | 0.4465 | -1.5851 | 1.9033 | -0.3141 | 0.4614 | -0.7837 | 2.0087 |
| | | MLE | -0.4934 | 0.4586 | -2.4012 | 3.4528 | -0.3723 | 0.5704 | -1.0886 | 3.5295 |
| | | MME | -0.7540 | 0.8926 | -3.3794 | 4.7901 | 1.2931 | 1.8546 | -0.9462 | 5.2136 |

**Figure 2.** $Bias, RMSE$ of $LSPF, ML, MM$ estimators for $\delta$ based on 100 simulations with $\delta = 0.5, 1, 2, 3, 4, 5$ and $n = 20$.

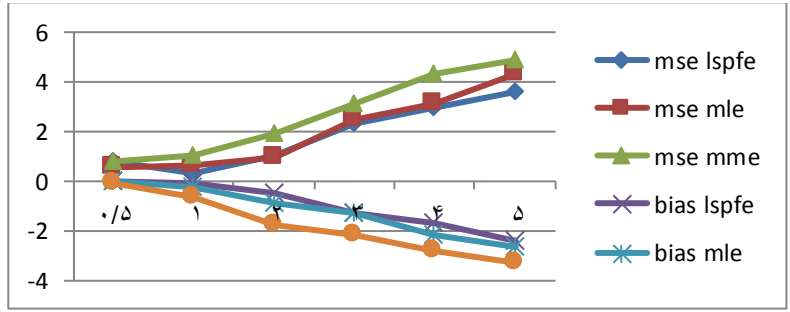

**Figure 3.** $Bias, RMSE$ of $LSPF, ML, MM$ estimators for $\delta$ based on 100 simulations with $\delta = 0.5, 1, 2, 3, 4, 5$ and $n = 50$.

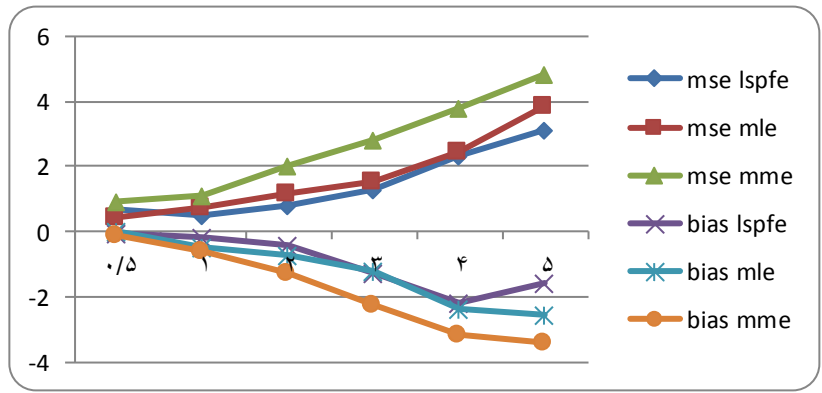

**Figure 4.** $Bias, RMSE$ of $LSPF, ML, MM$ estimators for $\delta$ based on 100 simulations with $\delta = 0.5, 1, 2, 3, 4, 5$ and $n = 100$.

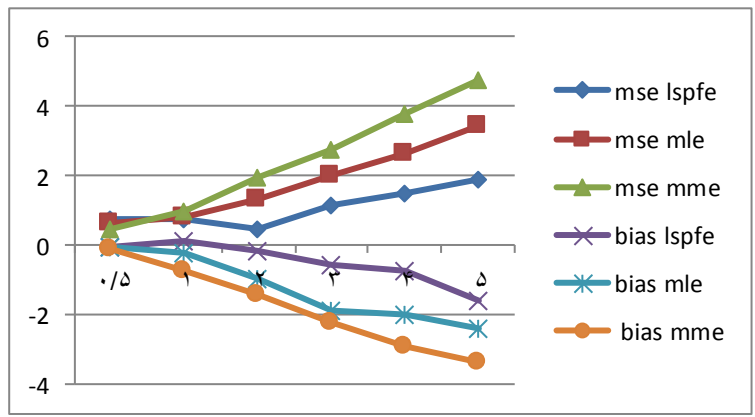

**Figure 5**. $Bias, RMSE$ of $LSPF, ML, MM$ estimators for $\beta$ based on 100 simulations with $\delta = 0.5, 1, 2, 3, 4, 5$ and $n = 20$.

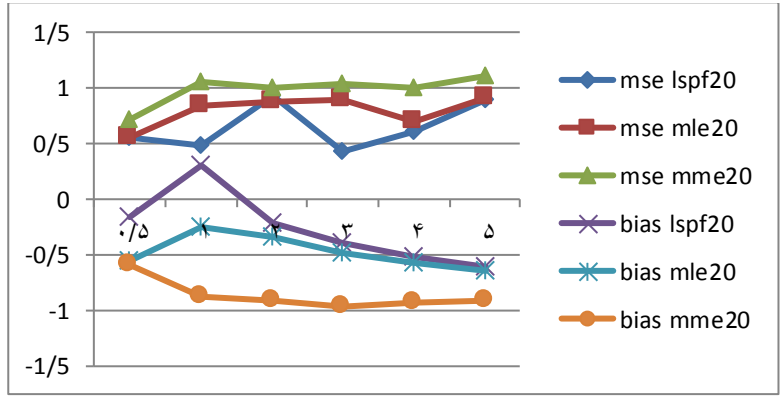

**Figure 6**. $Bias, RMSE$ of $LSPF, ML, MM$ estimators for $\beta$ based on 100 simulations with $\delta = 0.5, 1, 2, 3, 4, 5$ and $n = 50$.

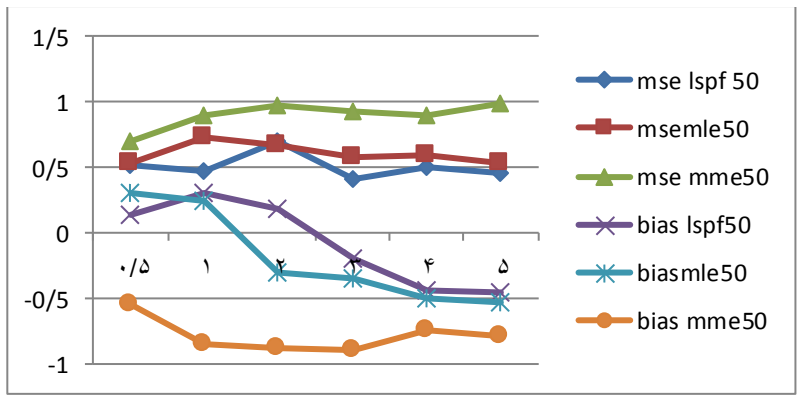

**Figure 7**. $Bias, RMSE$ of $LSPF, ML, MM$ estimators for $\beta$ based on 100 simulations with $\delta = 0.5, 1, 2, 3, 4, 5$ and $n = 100$.

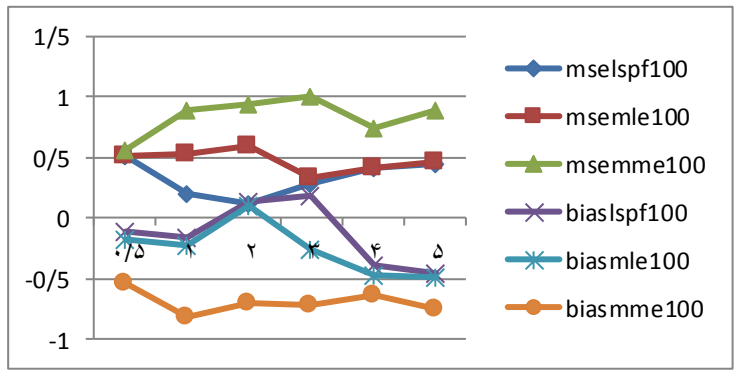

**Figure 8**. $Bias, RMSE$ of $LSPF, ML, MM$ estimators for $\gamma$ based on 100 simulations with $\delta = 0.5, 1, 2, 3, 4, 5$ and $n = 20$.

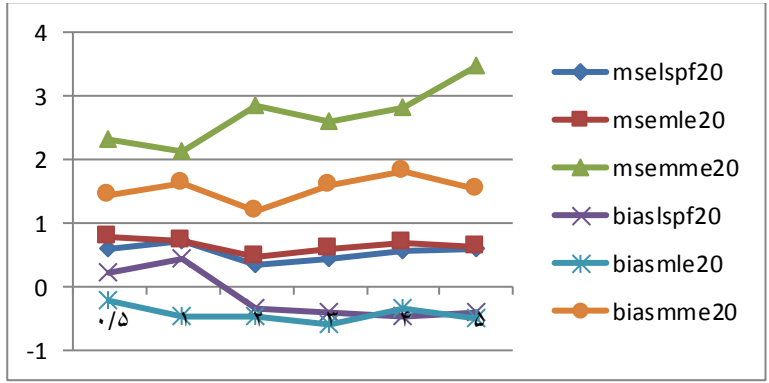

**Figure 9**. $Bias, RMSE$ of $LSPF, ML, MM$ estimators for $\gamma$ based on 100 simulations with $\delta = 0.5, 1, 2, 3, 4, 5$ and $n = 50$.

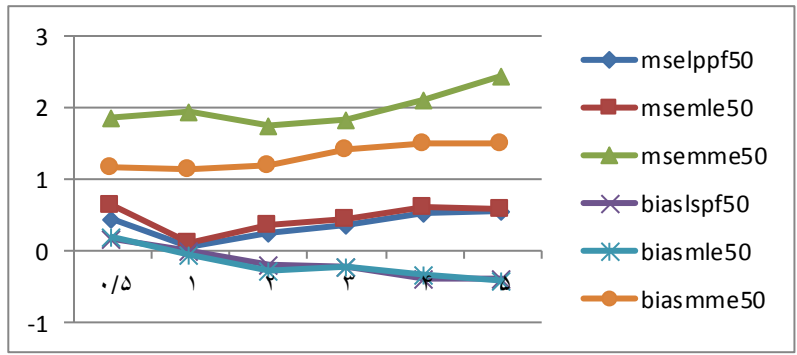

**Figure 10**. $Bias, RMSE$ of $LSPF, ML, MM$ estimators for $\gamma$ based on 100 simulations with $\delta = 0.5, 1, 2, 3, 4, 5$ and $n = 100$.

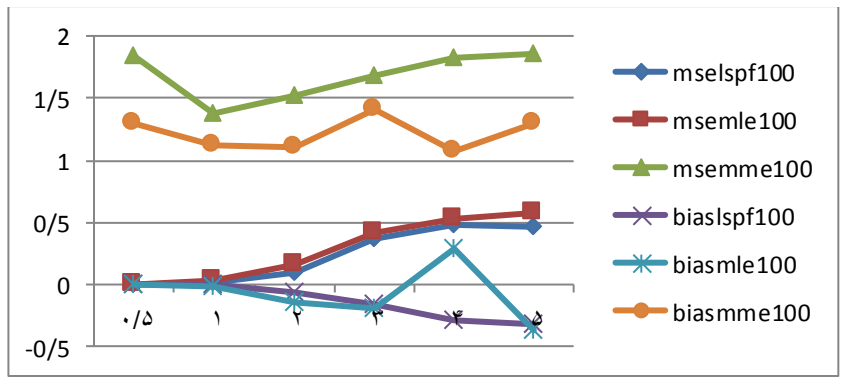

1. The *LSPFEs* have the less *RMSE* and absolute bias for every $n$ and $\delta$ constant.

2. If sample size will be large the *RMSE* and absolute bias for the *LSPFE* and *MLE* will be small.

3. If sample size will be large the *MME* has not very large variant, and this is almost independent to sample size.

4. When the $\delta$ is increase and the sample size is constant then the *RMSE* and absolute *Bias* are increase.

5. The *Bias* for *LSPFEs* are minus on the other hand *LSPFE* are less than real value , but the bias for *MLE* and *MME* are minus or plus.

## 5. Illustrate examples

We demonstrate the proposed method for the Three-parameter *RW* distribution in this section by using two data sets.one of them is with large sample size and the other is with small sample size.

### 5.2. Example 1

The first sample has been selected from Cohen [4] and Elderton and Johnson [5], is fitted an observed age distribution of holders of a certain type of life insurance policy. The data is in the table 7. Cohen [4]obtained the *MME* of this data, $\beta = 310.54659, \gamma = 339.7792125$ and $\delta = 40.043878$. We computed the *ML* and the *LSPF* estimation for $\beta, \gamma$ and $\delta$. The results are in table 8. Figures 11 and 12 show the density plots(fitted pdf versus empirical *PDF*) for the distribution plots (fitted *CDF* versus empirical *CDF*) for the three different estimation methods. The figures show that the *LSPF* estimators provide the best fit.

**Table 7**: The data for age of life insurance policy holders.

| Age in years | 5-14 | 15-24 | 25-34 | 35-44 | 45-54 | 55-64 | 65-74 | 75-84 | totals |
|---|---|---|---|---|---|---|---|---|---|
| frequencies | 1 | 56 | 167 | 98 | 34 | 9 | 2 | 1 | 368 |

**Table 8**:Estimates of the parameters for data I example 1.

|  | LSPFE | MLE | MME |
|---|---|---|---|
| $\delta$ | 8.58 | 15.88 | 40.04 |
| $\beta$ | 45.40 | 134.33 | 310.54 |
| $\gamma$ | 72.51 | 163.49 | 339.77 |

**Figure 11**: Fitted *PDF*s and the histogram for the three different estimation methods for data in example 1.

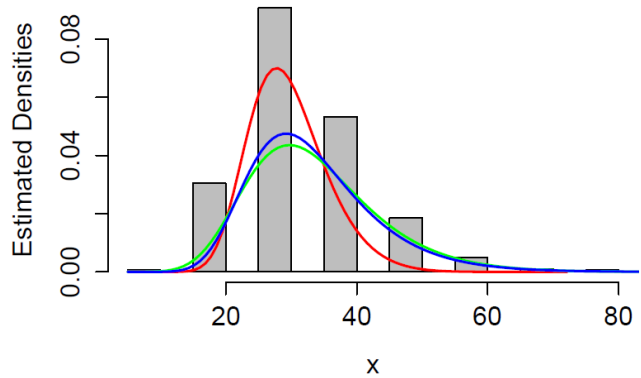

**Figure 12:** Fitted versus the empirical $CDF$ for the three different estimation methods for data in example 1.

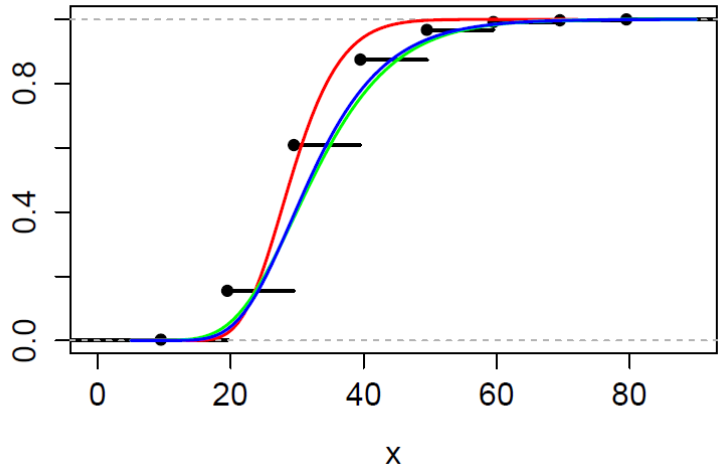

### 5.2. Example 2

We said if $X$ has the Weibull distribution then $-X$ has the $RW$ distribution. Next, we consider initially reported by Nagatsuka et al.[13].

This data has the three-parameter Weibull distribution and reporte in the table 9. We consider $-X$ that has the three-parameter $RW$ distribution.

This data is bearing´s fatigue life data. The $MLE$ and $LSPFE$ and $MME$ of parameters are in table 10.

Figures 13 and 14 show the density plots(fitted $PDF$ versus empirical $PDF$) for the distribution plots (fitted $CDF$ versus empirical $CDF$) for the three different estimation methods. The figures show that the $LSPF$ estimators provide the best fit.

**Table 9:** The data for age of life insurance policy holders.

-152.7 -172 -172.5 -173.3 -193 -204.7 -216.5 -234.9 -262.6 -422.6

Table 10: Estimates of the parameters for data in example 2.

|   | LSPFE | MLE | MME |
|---|---|---|---|
| $\delta$ | 0.7825 | 0.288 | 0.4037 |
| $\beta$ | 64.87 | 54.31 | 17.12 |
| $\gamma$ | -149.02 | -152.7 | 56.23 |

Figure 13: Fitted $PDF$s and the histogram for the three different estimation methods for data in example 2.

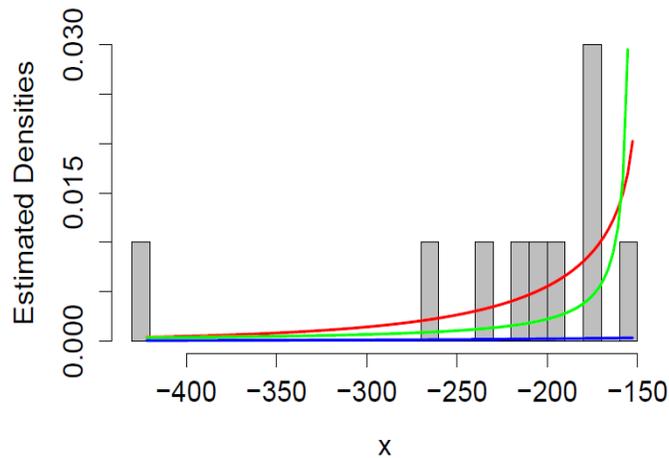

Figure 14: Fitted versus the empirical $CDF$ for the three different estimation methods for data in example 2.

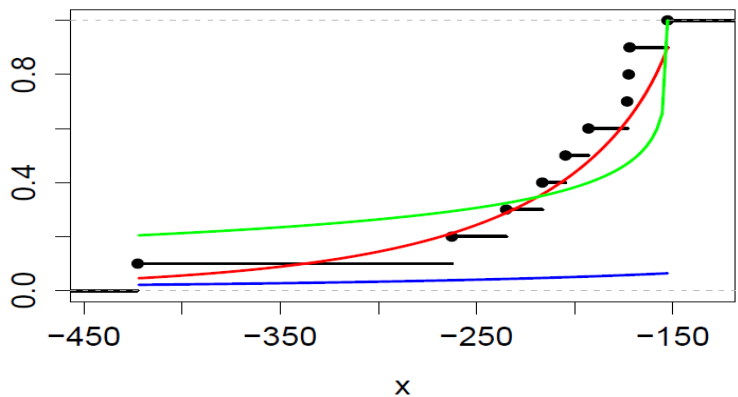

Examples show that $LSPFE$s are less $RMSE$ and bias then this is better than the $MLE$ and $MME$ for every sample size.

## 6. Concluding remarks

We consider three methods for estimation of parameters in the three-parameter *RW* distribution. The *MLE* and *MME* methods have studied in books and articles , but *LSPFE* has studied just for a little the three-parameter distributions, In article has proofed that *LSPFE*s provide the best fit.

## 7. Refrences